\documentclass[5p]{elsarticle}

\journal{Physica E}

\usepackage{color}
\usepackage{amsmath}
\usepackage{amssymb}

\begin{document}

\newcommand{\bn}{{\bf n}}
\newcommand{\bp}{{\bf p}}   
\newcommand{\br}{{\bf r}}
\newcommand{\bk}{{\bf k}}
\newcommand{\bv}{{\bf v}}
\newcommand{\brho}{{\bm{\rho}}}
\newcommand{\bj}{{\bf j}}
\newcommand{\wk}{\omega_{\bf k}}
\newcommand{\nk}{n_{\bf k}}
\newcommand{\eps}{\varepsilon}
\newcommand{\la}{\langle}
\newcommand{\ra}{\rangle}
\newcommand{\be}{\begin{equation}}
\newcommand{\ee}{\end{equation}}
\newcommand{\intl}{\int\limits_{-\infty}^{\infty}}
\newcommand{\dE}{\delta{\cal E}^{ext}}
\newcommand{\SE}{S_{\cal E}^{ext}}
\newcommand{\dsp}{\displaystyle}
\newcommand{\phit}{\varphi_{\tau}}
\newcommand{\p}{\varphi}
\newcommand{\cL}{{\cal L}}
\newcommand{\dphi}{\delta\varphi}
\newcommand{\dbj}{\delta{\bf j}}
\newcommand{\lra}{\leftrightarrow}
\newcommand{\comment}[1]{}

\allowdisplaybreaks 

\begin{frontmatter}

\title{The Boltzmann--Langevin approach: A simple quantum-mechanical derivation}

\author{K. E. Nagaev}

\address{Kotelnikov Institute of Radioengineering and Electronics, Mokhovaya 11-7, Moscow, 125009 Russia}

\begin{abstract}
We present a simple quantum-mechanical derivation of correlation function of Langevin sources
in the semiclassical Boltzmann--Langevin equation.  The specific case of electron--phonon scattering is 
considered. It is shown that the assumption of weak scattering leads to the Poisson nature of the
scattering fluxes.
\end{abstract}


\end{frontmatter}

\section{Introduction}

Nonequilibrium electrical noise in mesoscopic systems was always the subject of primary interest for Markus
B\"uttiker. The famous Landauer--B\"uttiker formula for the shot noise in quantum-coherent 
conductors became a cornerstone of modern theory of fluctuations~\cite{Buttiker92}. This formula was successfully applied 
to the calculations of shot noise in different mesoscopic systems with noninteracting electrons ranging from 
double-barrier resonant tunnel diodes~\cite{Buttiker91} to quantum-coherent metallic diffusive 
wires~\cite{Beenakker92}. However this method 
has difficulty in describing interacting electrons or systems with dephasing. To circumvent it, one has to 
introduce dephasing probes~\cite{Buttiker86,Buttiker88}, i.e. fictitious probes with voltages chosen such that they do not affect the electrical current but allow a replacement of quantum-coherent electrons in the conductor by electrons from reservoirs with a random phase. Yet the properties of the dephasing probes have to be somehow related to
the rate of actual microscopic scattering processes.

An alternative method for calculating the electrical noise in conductors in the limit of a large number of 
quantum channels is the Boltzmann--Langevin approach proposed by Kogan and Shul'man in 1969~\cite{Kogan69}. In this 
approach, the fluctuations of current and any other observable quantities are expressed in terms of the 
fluctuations of semiclassical distribution function, which obey the Boltzmann equation with a Langevin source 
in the right-hand side. This method appeared to be very efficient when calculating the hot-electron noise in diffusive metallic wires~\cite{Nagaev95,Kozub95}, frequency-dependent shot noise in metallic structures in the presence of external 
screening gates~\cite{Naveh97,Nagaev98}, and even the noise in hybrid superconductor -- normal-metal systems at voltages much higher
than the Thouless energy~\cite{Nagaev01}. More recently, it was extended to the case of spin-flip scattering
in ferromagnetic spin valves~\cite{Mishchenko03} and applied to Coulomb drag in clean double-layer 
systems~\cite{Chen15}

The key point in the Boltzmann -- Langevin approach is the derivation of the correlation function of Langevin sources. Kogan and Shul'man derived it assuming that the noise arises due to the randomness of electron scattering
by impurities and phonons. It was also assumed that all scattering events are independent, hence the scattering
of electrons
between a pair of states at a given space point presents a Poisson process, whose spectral density is proportional to its average rate.

Surprisingly, there were few attempts to derive the correlation function of Langevin sources directly from quantum-mechanical principles. In paper~\cite{Kogan91}, this correlation function was calculated using a sophisticated extension of Keldysh diagrammatic technique, which involved time-ordering on a four-branch temporal contour. The current paper presents a 
much simpler quantum-mechanical derivation of this quantity, which does not require a diagrammatic technique. 

\section{The general expression}

The standard semiclassical distribution function of electrons $n_{\bp}(\br t)$ presents the statistical average 
of the number of electrons in an element of phase space $\Delta p^3\times\Delta x^3$ 
divided by the number of quantum states in this element $\Delta N =\Delta p^3\,\Delta x^3/(2\pi\hbar)^3$,
which is centred at point $(\bp,\br)$. This implies that the statistical averaging is performed on top of the
coarse-grained averaging~\cite{Kirkwood46}.
Once the distribution function is known, one may easily calculate different measurable quantities like charge or current density as linear functionals of it. It can be shown in many different 
ways~\cite{Kirkwood46,Bogoliubov46,Keldysh64}
that this distribution function obeys the well-known Boltzmann equation
\be
 \frac{\partial\bar n_{\bp}}{\partial t}
 +
 \bv\,\frac{\partial\bar n_{\bp}}{\partial\br}
 +
 e{\bf E}\,\frac{\partial\bar n_{\bp}}{\partial\bp}
 =
 {I}_{col}\{\bar n_{\bp}\},
 \label{Boltz}
\ee
where $I_{col}$ is the collision integral that accounts for the electron scattering by impurities and phonons
or electron--electron scattering. 
This equation is valid provided that the semiclassical approximation holds. First, the characteristic length  of spatial variation of $n_{\bp}$ must be larger than the microscopic scale responsible for the scattering $\hbar/p_c$, where $p_c$ is the characteristic momentum of an electron. Second,  both the characteristic time of its variation and the inverse scattering rate must be larger than the characteristic time of the collision with impurity or phonon $\hbar/\eps_c$, where $\eps_c$ is the characteristic energy of an electron.


Along with the average distribution function, one may also be interested
in the correlation function of its fluctuations
$\la\delta n_{\bp_1}(\br_1t_1)\,\delta n_{\bp_2}(\br_2 t_2)\ra$, where the fluctuations 
$\delta n_{\bp_i} = n_{\bp_i} - \bar n_{\bp_i}$ are
only coarse-grained-averaged, and the statistical averaging is applied to their product~\cite{Kadomtsev57}. 
This quantity immediately 
gives the correlation functions of different observables. To calculate it, let us take a closer look at Eq.~\eqref{Boltz}. Apart from the time derivative, the terms in the left-hand side describe the deterministic motion of electrons
in the phase space due to smooth spatial variations of the distribution function and electrical potential. 
In contrast to this, the collision integral describes quantum-mechanical transitions of electrons between
the states with different momentum, which are assumed to be local in space and time. These transitions are random and should be considered as the source of noise if the semiclassical description is used.

As the structure of the Boltzmann equation without the drift terms resembles the equation of motion of
the Brownian particle, one may write the corresponding Lange\-vin equation for the distribution function. To this end, $\bar n_{\bp}$ should be replaced by $n_{\bp}$ and a random Langevin source $\delta J_{\bp}^{ext}(\br t)$ with zero average should be added~\cite{vanKampen} to the right-hand side of Eq.~\eqref{Boltz}. As the duration of an electron collision with an impurity or phonon is much smaller than characteristic time of variation of $n_{\bp}$, this source may be assumed to be delta correlated in time. 
Similarly, it should be delta correlated in space because of the local nature of the collisions. The momentum-dependent coefficient of these delta functions may be calculated as follows~\cite{Lax68}. Choose an interval of time $\Delta t$ much longer than the collision time $\hbar/\eps_c$ but so short that the distribution function cannot significantly change during this period. The direct integration of the Boltzmann -- Langevin equation over time gives the increment of $n_{\bp}$
\be
 \Delta n_{\bp} \equiv n_{\bp}(\br,t+\Delta t) - n_{\bp}(\br t) 
 = \int_0^{\Delta t} d\tau\,\delta J_{\bp}^{ext}(\br,t+\tau).
 \label{Dn-1}
\ee
Hence the correlation function of two such increments is given by a double integral
\begin{multline}
 \la\Delta n_{\bp_1}(\br_1)\,\Delta n_{\bp_2}(\br_2)\ra
 = \int_0^{\Delta t} d\tau_1 \int_0^{\Delta t} d\tau_2
 \\ \times
 \la\delta J_{\bp_1}^{ext}(\br_1,t+\tau_1)\,\delta J_{\bp_2}^{ext}(\br_2,t+\tau_2)\ra.
 \label{<DnDn>}
\end{multline}
The delta function of $\tau_1 - \tau_2$ eliminates one of the integrations, and the other 
reduces to a multiplication by $\Delta t$. Hence it follows from Eq.~\eqref{<DnDn>} that 
\begin{multline}
 \la\delta J_{\bp_1}^{ext}(\br_1,t_1)\,\delta J_{\bp_2}^{ext}(\br_2,t_2)\ra
 =
 \delta(t_1 - t_2)\,\delta(\br_1 - \br_2)
\\ \times
 \lim_{\Delta t\to 0} \frac{\la\Delta n_{\bp_1}(\br_1)\,\Delta n_{\bp_2}(\br_1)\ra}{\Delta t}.
 \label{<dJdJ>-def}
\end{multline}
This is our basic formula for calculating the correlation function of Langevin sources. 

\section{Equations of motion for the operators}

To carry
out the calculations to the end, we have to 
calculate the ratio in Eq.~\eqref{<dJdJ>-def} using quantum mechanics.
Consider the particular case of electron--phonon scattering in an elementary volume 
of size $\Delta x$ much smaller than the characteristic length at which the average distribution function or the 
electrical potential essentially change but much larger than $\hbar/p_c$. This allows us to describe the
scattering with a locally uniform Hamiltonian.
The Hamiltonian of the system $\hat H = \hat H_0 + \hat V$ is the sum of the noniteracting part
\be
 \hat H_0 = \sum_{\bp} \eps_{\bp}\, \hat a^{+}_{\bp} \hat a_{\bp} 
 + \sum_{\bk} \hbar\omega_{\bk}\, \hat b^{+}_{\bk} \hat b_{\bk}
 \label{H0}
\ee
and the part describing the electron--phonon interaction
\be
 \hat V = \sum_{\bp,\bk} (V_{\bk}\hat b_{\bk} + V_{-\bk}^{*}\hat b_{-\bk}^{+})\, 
 \hat a^{+}_{\bp+\hbar\bk} \hat a_{\bp},
 \label{V}
\ee
where $V_{\bk}$ are the matrix elements of electron--phonon interaction, and $\hat a_{\bp}$ and 
$\hat b_{\bk}$ are the annihilation operators for electrons and phonons, respectively. The time-dependent occupation-number operator $\hat n_{\bp} = \hat a^{+}_{\bp} \hat a_{\bp}$ obeys the Heisenberg 
equation~\cite{Shulman70}
\be
 \frac{d\hat n_{\bp}}{dt} = \frac{i}{\hbar}\,[\hat H, \hat n_{\bp}(t)].
 \label{Heisenberg}
\ee
As $[\hat n_{\bp}, \hat H_0]=0$, the time dependence of this operator is determined only by the weak 
electron--phonon coupling and may be considered as slow. Therefore Eq. \eqref{Heisenberg} may be solved
by iterations in $\hat V$. To this end, we perform a unitary transformation of all operators
\be
 \tilde A(t,\tau) = e^{-\frac{i}{\hbar}\hat H_0(t)\tau}\,\hat A(t)\,e^{\frac{i}{\hbar}\hat H_0(t)\tau},
 \label{unitary}
\ee
which brings Eq.~\eqref{Heisenberg} to the form
\be
 \frac{d\tilde n_{\bp}}{d\tau} = \frac{i}{\hbar}\,[\tilde V(t,\tau), \tilde n_{\bp}].
 \label{reduced}
\ee
If the time interval $\Delta t$ is much shorter than the relaxation time of the
distribution function due to the collisions, the increment of 
$\tilde n_{\bp}$ may be calculated to the second order in $\tilde V$, and then the
transformation inverse to Eq.~\eqref{unitary}  gives
\begin{multline}
 \Delta\hat n_{\bp} \equiv \hat n_{\bp}(t+\Delta t) - \hat n_{\bp}(t)
\\ =
 \frac{i}{\hbar}   \int_0^{\Delta t} d\tau\,[\tilde V(t,\tau-\Delta t),\, \hat n_{\bp}(t)]
 - \frac{1}{\hbar^2} \int_0^{\Delta t} d\tau' \int_0^{\tau'} d\tau''\,
\\ \times
    [\tilde V(t,\tau'-\Delta t),\, [\tilde V(t,\tau''-\Delta t),\, \hat n_{\bp}(t)]],
 \label{delta_n-1}
\end{multline}
where
\begin{multline}
 \tilde V(t,\tau) = 
  \sum_{\bp,\bk} \left(V_{\bk}e^{i\omega_{\bk}\tau}\hat b_{\bk} 
 + V_{-\bk}^{*}e^{-i\omega_{-\bk}\tau}\hat b_{-\bk}^{+}\right)\,
\\ \times
 e^{\frac{i}{\hbar}(\eps_{\bp}-\eps_{\bp+\hbar\bk})\tau}\, \hat a_{\bp+\hbar\bk}^{+} \hat a_{\bp}
 \label{V1}
\end{multline}
and all fermionic and bosonic operators are taken at time~$t$.
The density matrix of the system is assumed to be factorized into the electron and
phonon parts, which are diagonal in the same representation as $\hat H_0$. Therefore upon averaging 
Eq.~\eqref{delta_n-1}, the first summand vanishes and in the second summand, only diagonal terms are left.
The average products of four fermionic operators are decoupled into products of pair averages, e.g.
$\la\hat a^{+}_{\bp+\hbar\bk}\hat a_{\bp}\, \hat a^{+}_{{\bf q} - \hbar\bk} \hat a_{\bf q}\ra =
\delta_{\bp+\hbar\bk,\bf q}\,\hat n_{\bf q}(1-n_{\bp})$. As a result, all the arguments of
exponents are proportional to the difference $\tau' - \tau''$, and
\begin{multline}
 \la\Delta\hat n_{\bp}\ra =
 - \frac{2}{\hbar^2} \int_0^{\Delta t} d\tau' \int_0^{\tau'} d\tau''\sum_{\bk} 
\\
 \Bigl( \Bigl\{
  |V_{-\bk}|^2 \cos\!\left[
      (\hbar\omega_{-\bk}+ \eps_{\bp+\hbar\bk}-\eps_{\bp})(\tau'-\tau'')/\hbar
     \right](1+N_{-\bk})
\\ 
  +|V_{\bk}|^2 \cos\!\left[
   (\hbar\omega_{\bk}+\eps_{\bp}-\eps_{\bp+\hbar\bk})(\tau'-\tau'')/\hbar
  \right] N_{\bk} 
 \Bigr\} 
 \\ \times n_{\bp}\,(1-n_{\bp+\hbar\bk})
\\
 -\Bigl\{|V_{-\bk}|^2 
 \cos\!\left[
  (\hbar\omega_{-\bk}+\eps_{\bp+\hbar\bk}-\eps_p)(\tau'-\tau'')/\hbar
 \right] N_{-\bk}
\\
 +|V_{\bk}|^2 \cos\!\left[
   (\hbar\omega_{\bk}+\eps_{\bp}-\eps_{\bp+\hbar\bk})(\tau'-\tau'')/\hbar
  \right]\,(1+N_{\bk})\Bigr\}
 \\ \times
 n_{\bp+\hbar\bk}\,(1-n_{\bp}) \Bigr),
 \label{<dn>}
\end{multline}
where $N_{\bk} = \la b^{+}_{\bk} b_{\bk}\ra$ are the occupation numbers of phonon states.
The double integration of Eq.~\eqref{<dn>} over $\tau'$ and $\tau''$ can be replaced by a sequential
integration over $\tau'-\tau''$ and $\tau'$. If $\Delta t$ is much larger than the inverse 
characteristic frequencies in the arguments of the cosine functions, the integration over 
$\tau'-\tau''$ gives delta functions of its coefficients, and the subsequent integration over 
$\tau'$ reduces to a multiplication by $\Delta t$. Therefore
\be
 \frac{\la\Delta\hat n_{\bp}\ra}{\Delta t} =
 \sum_{\bk} [J(\bp+\hbar\bk \to \bp) - J(\bp \to \bp+\hbar\bk)],
 \label{dn/dt}
\ee
where
\begin{multline}
 J(\bp \to \bp+\hbar\bk) =
 \frac{2\pi}{\hbar}
 \Bigl[ |V_{-\bk}|^2\,\delta(\hbar\omega_{-\bk}+\eps_{\bp+\hbar\bk}-\eps_{\bp})\,
\\ \times (1+N_{-\bk})
        +|V_{\bk}|^2\,   \delta(\hbar\omega_{\bk}+\eps_{\bp}-\eps_{\bp+\hbar\bk})\,N_{\bk}
 \Bigr]\, 
\\ \times
 n_{\bp}\,(1-n_{\bp+\hbar\bk})
 \label{J}
\end{multline}
is the scattering flux {} from state $\bp$ to state $\bp + \hbar\bk$. The right-hand side of
Eq.~\eqref{dn/dt} is exactly the
electron--phonon collision integral $I_{col}$ that appears in Eq.~\eqref{Boltz}. 

\begin{figure}
  \centering
  \includegraphics[width=0.7\columnwidth]{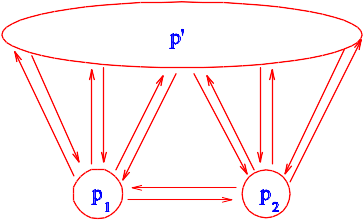}
  \caption{(color online). The scattering fluxes that appear in Eq.~\eqref{<dJdJ>-2}. States
  $\bp_1$ and $\bp_2$ exchange electrons with each other and with the rest of states $\bp'$.}
  \label{fig1}
\end{figure}

Now we calculate the average product of two increments $\Delta\hat n_{\bp_1}$ and $\Delta\hat n_{\bp_2}$ 
with the same $t$ and $\Delta t$. To the second order in the interaction potential, 
\begin{multline}
 \la\Delta n_{\bp_1}\,\Delta n_{\bp_2}\ra = 
 -\frac{1}{\hbar^2} \int_0^{\Delta t} d\tau_1\int_0^{\Delta t} d\tau_2
\\ \times
 \la[\tilde V(t,\tau_1-\Delta t),\, \hat n_{\bp_1}(t)]\,[\tilde V(t,\tau_2-\Delta t),\, \hat n_{\bp_2}(t)]\ra.
 \label{<n1n1>}
\end{multline}
The averaging is performed according to the same rules as in Eq.~\eqref{delta_n-1}, and the arguments of complex
exponents appear to be proportional to the difference $\tau_1 - \tau_2$.
Much like when calculating the collision integral, we replace the double integration over $\tau_1$ and $\tau_2$
by a sequential integration over $\tau_1-\tau_2$ and over $\tau_1$. The former integration transforms the exponents in (\ref{<n1n1>}) into delta functions of coefficients of $\tau_1-\tau_2$, and the latter reduces to multiplication by
$\Delta t$. Therefore one obtains
\begin{multline}
 \frac{\la\Delta n_{\bp_1}\,\Delta n_{\bp_2}\ra}{\Delta t} =
 \delta_{\bp_1\bp_2} \sum_{\bk} [J(\bp_1 \to \bp_1+\hbar\bk) \\+ J(\bp_1+\hbar\bk \to \bp_1]
 - J(\bp_1 \to \bp_2) - J(\bp_2 \to \bp_1),
 \label{<dJdJ>-2}
\end{multline}
where the scattering fluxes $J$ are given by Eq.~\eqref{J}. The substitution of this ratio into 
Eq.~\eqref{<dJdJ>-def} gives precisely the same expression for the correlation function of Langevin sources
that was originally obtained in Ref.~\cite{Kogan69}. The full expression for it is given in \ref{app:full} by 
Eq.~\eqref{<n1n1>-2}. The physical interpretation of this expression was discussed in the above paper and
is illustrated in Fig.~\ref{fig1}. The arrows show the scattering fluxes between different pairs of local states, which are independent of each other. If states $\bp_1$ and $\bp_2$ coincide, all the outgoing and incoming scattering fluxes are correlated with themselves and give contributions proportional to their average rates. If these states are different, the contribution is given only by $J(\bp_1 \to \bp_2)$ and $J(\bp_2 \to \bp_1)$. Our calculations show
that the independence of different scattering fluxes and their Poisson statistics are the direct consequences of the
assumption of weak interaction of electrons with the scatterers.

The Boltzmann equation Eq.~\eqref{Boltz} is nonlinear because of the electron--phonon collision integral \eqref{dn/dt}. However in actual problems, the fluctuations of the distribution function are effectively averaged over macroscopic lengths much larger than $\Delta x$ 
and therefore are small. Hence the Boltzmann - Langevin equation may be obtained from Eq.~\eqref{Boltz} by linearizing it with respect to the fluctuation $\delta n_{\bp}$ and adding the Langevin source to the right-hand 
side
\be
 \frac{\partial\delta n_{\bp}}{\partial t}
 +
 \bv\,\frac{\partial\delta n_{\bp}}{\partial\br}
 +
 e{\bf E}\,\frac{\partial\delta n_{\bp}}{\partial\bp}
 =
 \delta{I}_{e-ph} + \delta J_{\bp}^{ext}.
 \label{BL}
\ee
Together with the expression for the correlation function of Langevin sources given by Eqs. \eqref{<dJdJ>-def}
and \eqref{<dJdJ>-2}, it presents the full set of equations for calculating the noise.

\section{Conclusions}

We have derived the correlation function of Langevin sources in the Boltzmann -- Langevin equation
for the fluctuations of the semiclassical distribution function. It was postulated that the sources
are delta correlated in space and time, but the momentum-dependent part was rigorously calculated using quantum-mechanical
equation of motion for the operators of occupation numbers. If the electron--phonon interaction is assumed
to be weak, the scattering fluxes between different pairs of electron states may be considered 
as independent Poisson processes, and the correlation function of Langevin sources is the sum of their
correlation functions taken with the appropriate signs. This coincides with the conditions for the validity
of the standard collision integral in the Boltzmann equation.

\section*{Acknowledgments}

I am grateful to A.~Ya.~Shul'man for reading the ma\-nu\-script of the paper and a useful discussion.

\appendix
\section{\label{app:full}}


The full correlation function of Langevin sources for the case of electron--phonon scattering
is given by the following expression:
\begin{multline}
  \la\delta J_{\bp_1}^{ext}(\br_1,t_1)\,\delta J_{\bp_2}^{ext}(\br_2,t_2)\ra
 =
\delta(t_1 - t_2)\,\delta(\br_1 - \br_2)
\\ \times 
 \frac{2\pi}{\hbar}\sum_{\bk}
 \Biggl\{ 
  \delta_{\bp_1,\bp_2}
  \Bigl[
  |V_{\bk}|^2\, \delta(\eps_{\bp_1+\hbar\bk} - \eps_{\bp_1} - \hbar\omega_{\bk})\,(1 + N_{\bk})
\\ +
  |V_{-\bk}|^2\, \delta(\eps_{\bp_1+\hbar\bk} - \eps_{\bp_1} + \hbar\omega_{-\bk})\,N_{-\bk}
  \Bigr]
\\ \times
  n_{\bp_1+\hbar\bk}\,(1 - n_{\bp_1})
\\ +
  \delta_{\bp_1,\bp_2}
  \Bigl[
  |V_{-\bk}|^2 \delta(\eps_{\bp_1} - \eps_{\bp_1+\hbar\bk} - \hbar\omega_{-\bk})\,(1 + N_{-\bk})
\\ +
  |V_{\bk}|^2 \delta(\eps_{\bp_1} - \eps_{\bp_1+\bk} + \hbar\omega_\bk)\,N_{\bk}
  \Bigr]
\\ \times
  n_{\bp_1}\,(1 - n_{\bp_1+\hbar\bk_1})
\\ -
  \delta_{\hbar\bk,\bp_2-\bp_1}
  \Bigl[
   |V_{\bk}|^2\, \delta(\eps_{\bp_2} - \eps_{\bp_1} - \hbar\omega_{\bk})\,(1 + N_{\bk})
\\ +
   |V_{-\bk}|^2\, \delta(\hbar\omega_{-\bk} + \eps_{\bp_2} - \eps_{\bp_1})\,N_{-\bk}
  \Bigr] 
\\ \times
  (1 - n_{\bp_1})\,n_{\bp_2}  
\\   -
  \delta_{\bk,\bp_1-\bp_2}
  \Bigl[
   |V_{\bk}|^2\, \delta(\eps_{\bp_1} - \eps_{\bp_2} - \hbar\omega_{\bk})\,(1 + N_{\bk})
\\
   +
   |V_{-\bk}|^2\, \delta(\hbar\omega_{-\bk} + \eps_{\bp_1} - \eps_{\bp_2})\,N_{-\bk}   
  \Bigr]
\\ \times
   n_{\bp_1}\,(1 - n_{\bp_2})
 \Biggl\}.
\label{<n1n1>-2}
\end{multline}
%


\end{document}